\documentclass[10pt,leqno]{amsart}
\usepackage{graphicx}
\usepackage{indentfirst,csquotes}

\usepackage{amssymb,amsthm,amsmath}
\usepackage{xcolor,paralist,hyperref,titlesec,fancyhdr,etoolbox}

\usepackage[backend=bibtex,style=authoryear]{biblatex}
\usepackage{times}
\usepackage{helvet}
\usepackage{courier}
\usepackage{graphicx}
\usepackage{caption}
\usepackage{url}
\usepackage{multirow}
\usepackage{multicol}
\usepackage{hyperref}

\titleformat{\section}[display]{\normalfont\huge\bfseries\centering}{}{0pt}{\Large}
\titleformat{\subsection}{\normalfont\large\bfseries}{\thesubsection}{1em}{}
\titlespacing*{\subsection}{0pt}{0.5ex plus 1ex minus .2ex}{0.5ex plus .2ex}

\hypersetup{ colorlinks=true, linkcolor=black, filecolor=black, urlcolor=black }

\usepackage{lipsum}

\begin{document}

\title{Analysis of Propaganda in Tweets From Politically Biased Sources}

\author{Vivek Sharma\textsuperscript{1}, Mohammad Mahdi Shokri\textsuperscript{1}, Sarah Ita Levitan\textsuperscript{2,1}, Elena Filatova\textsuperscript{3,1}, Shweta Jain\textsuperscript{4,1}\\
\noindent\textsuperscript{1}The Graduate Center, CUNY \texttt{\{vsharma,mshokri\}@gradcenter.cuny.edu}\\
\textsuperscript{2}Hunter College, CUNY \texttt{sarah.levitan@hunter.cuny.edu}\\
\textsuperscript{3}New York City College of Technology, CUNY \texttt{efilatova@citytech.cuny.edu}\\
\textsuperscript{4}John Jay College of Criminal Justice, CUNY \texttt{sjain@jjay.cuny.edu}\\
}

\begin{abstract}
News outlets are well known to have political associations, and many national outlets cultivate political biases to cater to different audiences. Journalists working for these news outlets have a big impact on the stories they cover. In this work, we present a methodology to analyze the role of journalists, affiliated with popular news outlets, in propagating their bias using some form of propaganda-like language. We introduce JMBX(Journalist Media Bias on X), a systematically collected and annotated dataset of 1874 tweets from Twitter (now known as X). These tweets are authored by popular journalists from 10 news outlets whose political biases range from extreme left to extreme right. We extract several insights from the data and conclude that journalists who are affiliated with outlets with extreme biases are more likely to use propaganda-like language in their writings compared to those who are affiliated with outlets with mild political leans. We compare eight different Large Language Models  (LLM) by OpenAI and Google. We find that LLMs generally performs better when detecting propaganda in social media and news article compared to BERT-based model which is fine-tuned for propaganda detection. While the performance improvements of using large language models (LLMs) are significant, they come at a notable monetary and environmental cost. This study provides an analysis of both the financial costs, based on token usage, and the environmental impact, utilizing tools that estimate carbon emissions associated with LLM operations.
\end{abstract}

\maketitle

\bigskip

\section{Introduction}

As defined by Barron et al.\cite{barron2019proppy} and Da San Martino et al.~\cite{da2020prta}, propaganda involves the deliberate expression of opinion with the intent to influence the opinion or to invoke some action. In recent years, the use of social media platforms to spread propaganda and misinformation has become increasingly prevalent. According to a 2021 Statista\cite{statista21}, 
identified state actors spreading propaganda and misinformation increased to 81 from 28 in 2017. Propaganda in social media is not limited to state actors. In fact, news outlets often utilize some form of propaganda to reach their audience by slanting their report toward the consumers' expectations, also known as media bias~\cite{gentzkow2006media}. A study by Baron ~\cite{baron2006persistent} shows that there is an established arrangement between the media house and the journalist. The study presents a theory that journalists can deliberately express biased opinions in their stories to advance their career prospects and news outlets choose to be biased to boost their profits. 
Paul et al.~\cite{paul2006detect} point out that it is often not up to journalists to determine what their reader wants to read. Instead, it is the public who wants their belief extolled and confirmed. Media bias affects public opinion causing polarization.

News outlets and the journalists associated with them often interact with the public through social media. The micro-blogging platform, X (formerly Twitter), is a popular medium of such communication which helps news outlets and their affiliates engage directly with their readers. According to a survey conducted in 2021 by Pew Research~\cite{pew22} seven in ten journalists prefer Twitter for their work-related communications and 79\% of them believe that social media help them to better engage with their audience. A study of 22,000 tweets conducted in 2011 by Lasorsa et al.~\cite{lasorsa2012normalizing} shows that journalists freely express their opinions on the X microblogging platform. This study highlights that engagement with social media followers promotes ``end-user journalism''. The extent of the end-user journalism is evident from the results presented in the study by Noguera et al~\cite{noguera2013open}. This study points out that 5\% of the tweets directly request information from the followers, 27\% are direct replies to the followers and 32\% of the tweets contain links to sources that are more often external to the news outlet. 

In Section 3 we introduce a publicly available dataset, JMBX\footnote{\url{https://figshare.com/s/349c826391f77c0a4899}}, that contains 1,874 annotated tweets from several news outlets. These tweets are labeled as either ``propaganda'' (containing a certain type of propaganda technique) or ``non propaganda''. Given this annotation, we explore the relationship between propaganda and the bias of the affiliated news outlet.

In Section 4 we present experiments using a fine-tuned BERT model and eight different Large Language Models (LLMs) on the task of propaganda detection. Our findings indicate that LLMs outperform the BERT model, with further improvements observed when employing the Chain of Thought prompting technique ~\cite{wei2022chain} in the data set. 

In Sections 5, we provide an estimate of environmental impact in running LLMs for propaganda detection offering a holistic evaluation of the study's implications.

Thus, we introduce a novel dataset for propaganda detection and answer the following research questions.

    \textbf{RQ1:} How frequently do journalists from news outlets with varying degrees of known biases (left, right, or center) exhibit characteristics indicative of propaganda in their tweets, and are those affiliated with organizations that have extreme biases more likely to use propaganda compared to their counterparts in moderately biased organizations?
    
    \textbf{RQ2:} Can Large Language Models (LLM) be used to detect propaganda in text? How does the performance of LLM compare with the performance of BERT model fine-tuned to detect propaganda. Furthermore, do prompt engineering techniques help improve performance of LLM in detecting propaganda?
    
    \textbf{RQ3:} What is the environmental cost of using LLMs to detect propaganda?

\begin{figure*}[htp]
    \centering
    \includegraphics[width=16cm]{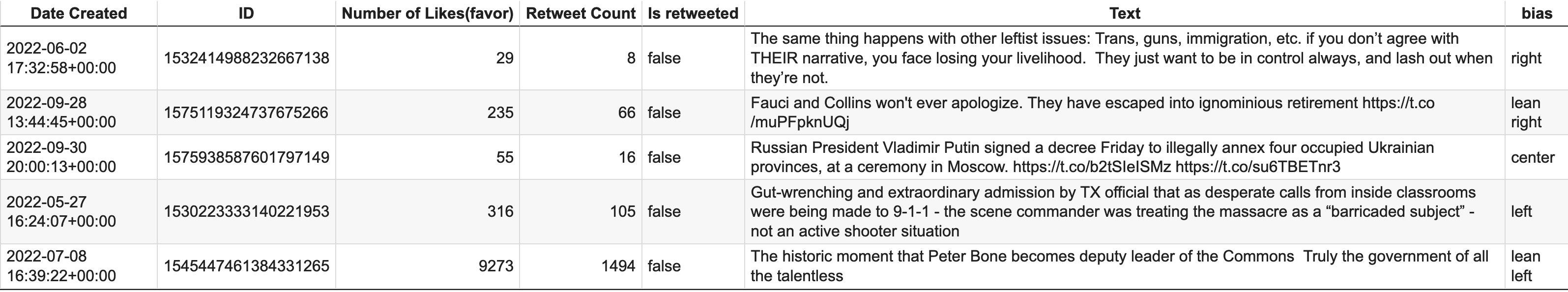}
    \caption{A sample of the JMBX dataset.}
    \label{fig:sample_bias_dataset}
\end{figure*}

\section{Theoretical Background}

In 1939, social scientists Alfred and Elizabeth Lee described seven propaganda techniques in their book ``The Fine Art of Propaganda''~\cite{lee1939fine}. More recently, Da San Martino et al.~\cite{martino2020survey} defines 18 propaganda techniques that are found in digital media. The most frequently used propaganda techniques are: loaded language, name calling, repetition, and slogans~\cite{martino2020semeval}. Other propaganda techniques include: doubt, appeal to fear, flag-waving, bandwagon, reduction ad hitlerum, causal oversimplification, black-and-white fallacy, whataboutism, thought terminating cliches, and, exaggeration or minimization. According to Huang et al.~\cite{huang2022faking}, people often use appeal to authority as a propaganda technique. Huang et al. use this technique in their generative AI model which generates synthetic texts closely resembling examples of human-written disinformation.
Among the less frequently used propaganda techniques are: strawman, red herring, and obfuscation~\cite{yu2021interpretable}.

Researchers have created several datasets to study text-based propaganda. These datasets contain news articles, social media blogs and limited-length microblogs. Some datasets are topical such as the COVID-19 pandemic~\cite{naseem2021covidsenti}~\cite{memon2020characterizing}, the UK General elections~\cite{nizzoli2021coordinated}, and the Russia-Ukraine war~\cite{haq2022twitter}.

PTC (Propaganda Technique Corpus)~\cite{da2019fine}
dataset contains 550 articles annotated with 18 propaganda techniques. These articles are extracted from news sources of various political bias determined by Media Bias Fact Check (MBFC), an independent website that relies on human evaluators and a methodical approach~\cite{mbfc_method} to determine the bias of media sources. This data set is suitable for training and testing natural language processing models that perform both fragment and sentence level classifications as well as span identification and identifying the propaganda technique.
    
QCRI's propaganda corpus ~\cite{barron2019proppy} contains 51.3K  articles from 104 news sources that appeared between October 2017 and December 2018. The articles are classified as propaganda or trustworthy. All articles from a news outlet are labeled based on their bias level (left, right, center) reported on MBFC. A total of 94 news sources are considered as sources for trustworthy articles while 10 sources are used for articles that contained propaganda. The following information is recorded for each article: title, text, average sentiment, publication date, and official source name. Geographical information is added to the dataset with the help of data made available by the Global Database of Events, Language and Tone Project (GDELT)~\cite{leetaru2013gdelt}.

The TWEETSPIN dataset~\cite{vijayaraghavan2022tweetspin} contains 210,392 tweets in the English language. All tweets in the dataset have at least one reply calling out the tweet with propaganda along with the technique of propaganda used. Currently, this dataset is not publicly available.

Recently researchers explore the use of LLMs as a tool to classify text and draw insights. LLMs use large amounts of training data and the power of transformers to learn relations between sentences and predict the next sentence, generating highly accurate pieces of text. Consequently, researchers experiment with using LLMs to detect bias in text~\cite{lin2024investigating,fan2024biasalert}, identifying subjective language~\cite{shokri-etal-2024-subjectivity,suwaileh2024thatiar}, and detecting fake news~\cite{liu2024detect}. Another research direction deals with various prompt engineering techniques~\cite{wei2022chain,zhang2022automatic,brown2020language} and evaluation of the performance of LLMs as a classification tool compared to traditional Natural Language Models such as BERT. This is a paradigm shift from using models trained to perform a specialized task to using a generalized model in a variety of classification tasks.
Jones et al.~\cite{jones2024detecting} used GPT 3.5 turbo to detect propaganda in SemEval-2020 Task 11 dataset, which is also known as PTC dataset. Sprenkamp~\cite{sprenkamp2023large} extend this approach by experimenting with five different GPT-3 and GPT-4 models. However, these studies are limited to news articles and use an LLM from a single provider. With the introduction of Google's Gemini models, competition in the LLM market has increased. Therefore, it is worthwhile to compare the performance of LLMs against each other. In our study, we introduce a new social media sourced dataset JMBX and evaluate eight LLM models, including five from OpenAI’s GPT series and three from Google's Gemini. To the best of our knowledge, this is the largest known comparison of models in this domain. We measure the performance of these models in detecting propaganda in news articles and social media microblogs. Additionally, we provide an estimate of the carbon footprint of any product that could be based on this research.

\section{Overview of the JMBX Dataset}

The microblogging site X, formerly known as Twitter, provides access to APIs that researchers can use to curate tweets and their replies for social media analysis. In this paper, our goal is to demonstrate a methodical approach to extract tweets by journalists who are affiliated with news organizations that are known to have certain political biases. Using this dataset we analyze the prevalence of propaganda language in the tweets posted by authors affiliated with biased news outlets.

The dataset consists of 1874 annotated tweets from journalists affiliated with 10 news outlets. The data was downloaded from Twitter between September 27, 2022 and October 3, 2022, using the Twitter Streaming API. Each record contains the tweet id, number of likes, retweet count, retweeted status. The last column contains the label, which is the propaganda technique used in the tweet. The tweet text is redacted to comply with twitter's data sharing policy. In addition to this, each record has a bias column with values left, right, lean right, lean left, and center labeled through distant supervision as per the ratings on AllSides Media Bias.\footnote{\url{https://www.allsides.com/media-bias/ratings}}

\subsection{Data Collection Methodology}

\begin{figure}[t]
    \centering
    \includegraphics[width=\columnwidth]{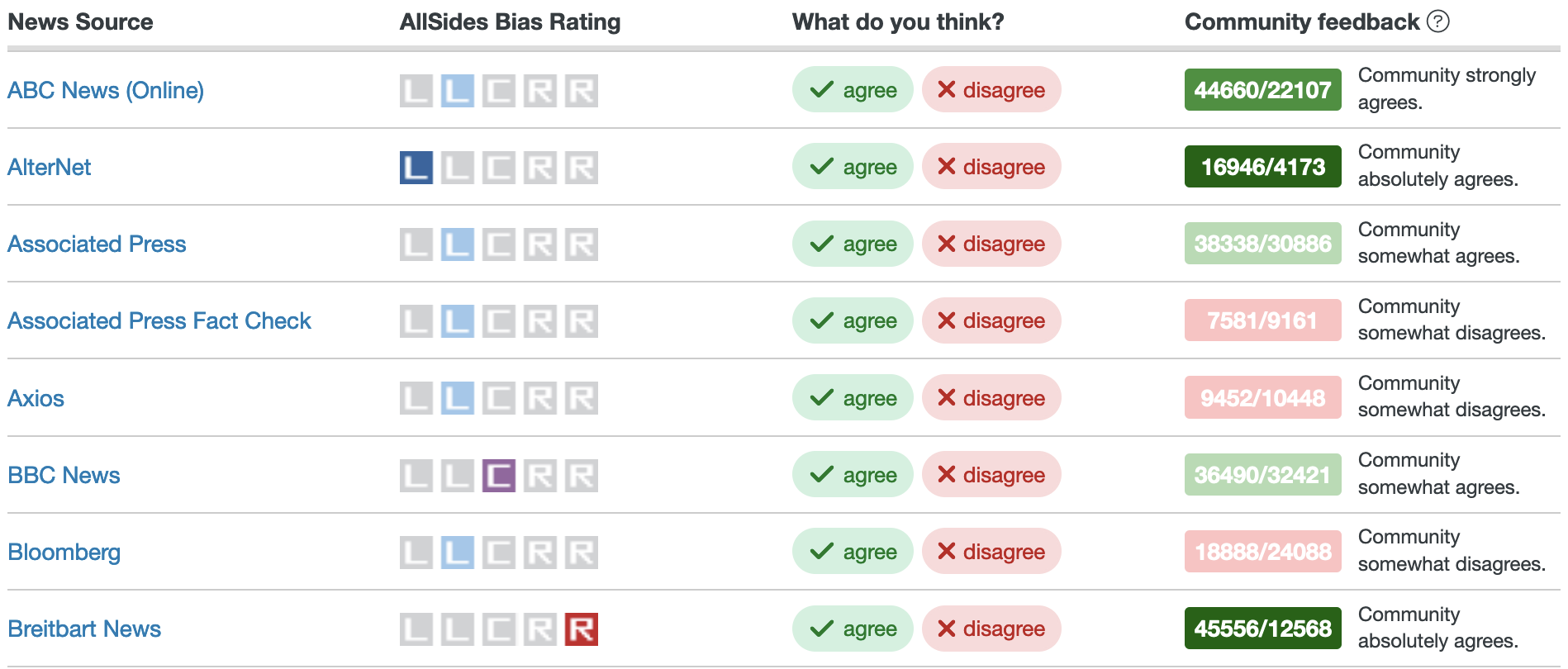}
    \caption{Sample ratings from Allsides media bias.}
    \label{fig:news_outlet_bias}
\end{figure}

AllSides Media Bias is an independent, multi-partisan agency that uses multi-partisan editorial reviews by trained experts and Blind Bias Surveys by readers to assign political lean (left, lean left, center, lean right, right) to news outlets. AllSides describes a transparent mechanism for generating these ratings which incorporates community feedback where the general public can agree or disagree with the ratings. While the community feedback does not change the original bias rating, it helps to confirm or refute the ratings based on public opinion.  The community feedback is a 8-point Likert scale from  ``Absolutely agree'' to ``Absolutely Disagree''. Figure~\ref{fig:news_outlet_bias} shows a sample screenshot. 

We used the bias ratings from AllSides to begin collecting our data. First, we select news outlets whose bias ratings received a majority of ``Absolutely agree'', ``Strongly agree'' and ``Agree'' ratings from the public. Next, we select news outlets with the highest number of followers on Twitter as this number is a well known measure of influence in a social network~\cite{kim2020value}. The following 10 news outlets are the result of the above filtering procedure. The results are as of September 2022 from each bias category, as reported on AllSides Media Bias Rating\textsuperscript{TM}.

\begin{itemize}
    \item \textbf{Left:} MSNBC, The New Yorker
    \item \textbf{Lean left:} ABC News, The Guardian
    \item\textbf{Center:} Forbes, Reuters
    \item\textbf{Lean right:} The New York Post, The Epoch Times
    \item\textbf{Right:} Breibart News, The Daily Wire
\end{itemize}

We performed a keyword search to find the profiles of journalists affiliated with each outlet by using the news outlet’s official name as the keyword. In this work, we assume that a journalist associated with a particular news organization is aligned with the political beliefs of that organization. The criteria used to select the journalists associated with the aforementioned news outlets are:
\begin{itemize}
    \item{The profile must have had some activity in the past three months.}
    \item{The journalist’s bio must contain the name of the news outlet they are associated with.}
    \item{Journalists who covered political news are prioritized over sports, lifestyle, and travel journalists.}
    \item{The journalist have the Twitter legacy verified checkmark at the time of data collection. (This data was collected before the checkmark's became available for purchase.)}
\end{itemize}

Using the above methodology, five journalists from each selected news outlet, who have the highest number of followers among their peers affiliated with the same outlet, are chosen with the exception of the news outlets considered centrist i.e., Forbes and Reuters. Twitter searches for journalists affiliated with Forbes and Reuters do not provide any results that match the above criteria. Therefore, instead of individual journalists, the official Twitter handle of Reuters and Forbes are used to collect tweets and are labeled as political center. We use Twitter Streaming API to obtain 1,500 most recent tweets by each journalist. For Reuters and Forbes, we collect 5,000 most recent tweets from the official Twitter handles. Data is cleaned by removing URLs, mention (@) symbols, and some tweets that abruptly ended with ``...''. To focus solely on text, we remove emojis.

\subsection{Annotation of the JMBX Dataset}
\label{sec:annotation}

We use the annotation services provided by A Data Pro~\footnote{\url{https://adata.pro}}, the organization that annotated the SemEval 2020 dataset which is widely used by researchers in the domain of text-based propaganda detection. The dataset comprises a balanced set of 2000 tweets, each with an equal number of instances featuring positive and negative sentiment scores. The annotation process follows the guidelines provided by Martino et al.~\cite{martino2020semeval}. Two annotators and one consolidator, who serves as the subject matter expert, are employed to classify the dataset at a fine-grained level, providing each tweet with a label from the 18 propaganda techniques~\cite{martino2020semeval}.
If any strategy from the list predefined propaganda strategies is detected in a tweet then the annotators label the tweet as \emph{propaganda}, otherwise it is labeled \emph{non-propaganda}. If the two independently and asynchronously-working annotators agree, then their label is used as the final label for the tweet. If the two annotators disagree then the label provided by the consolidator is used as the final label. The inter-rater agreement on this labeling task is found to be substantial, with a Cohen’s Kappa coefficient of 0.79. To resolve any discrepancies, a consolidator reviewed the annotations and facilitated discussions with the annotators to achieve consensus on all entries. After removing duplicates and incomplete or meaningless entries, the final annotated dataset comprised 1,874 tweets.

\subsection{Initial Insights from the JMBX Dataset}

\begin{figure}[t]
    \centering
    \includegraphics[width=\columnwidth]{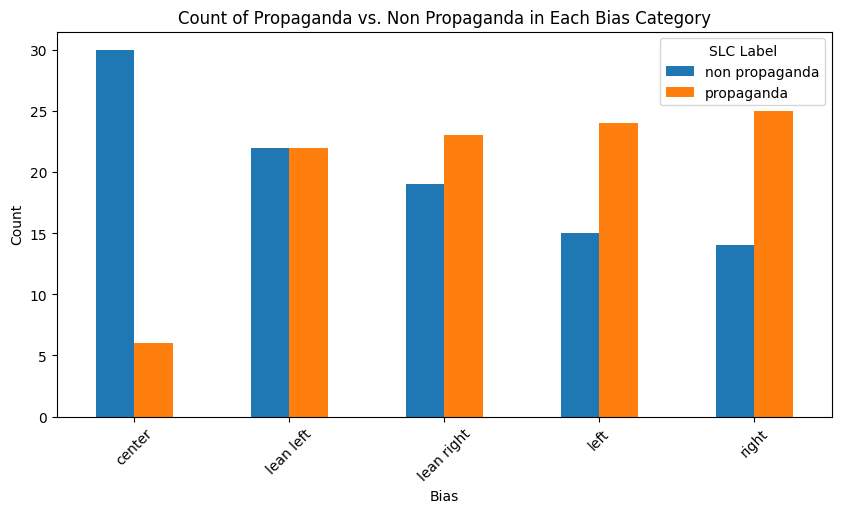}
    \caption{Propaganda in each bias category}
    \label{fig:bias_prop}
\end{figure}

Figure~\ref{fig:bias_prop} demonstrates that tweets labeled as non-propaganda are most prevalent in the center bias category, which aligns with expectations given that many of these tweets originate from centrist outlets. In contrast, tweets from sources with extreme bias exhibit a higher proportion of propaganda as compared to non-propaganda.

\begin{figure*}[ht]
    \centering
    \includegraphics[width=16cm]{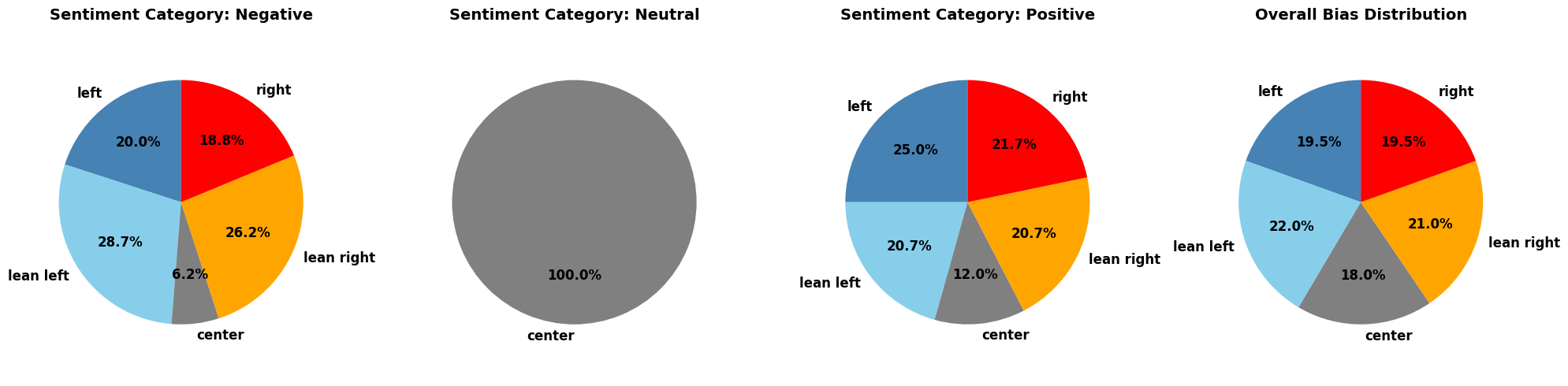}
    \caption{Sentiments vs Political Bias in JMBX dataset}
    \label{fig:sentiment_bias_4}
\end{figure*}

We conduct a sentiment analysis of the tweets using the TextBlob package, which assigns each tweet a continuous sentiment score ranging from -1 to 1. Tweets with negative sentiment are assigned values between -1 and -0.33, while positive sentiment tweets are assigned values between 0.34 and 1. Neutral sentiment is defined by values between -0.32 and 0.33. Figure~\ref{fig:sentiment_bias_4} illustrates the percentage distribution of tweets across different sentiment values (negative, neutral, positive) for different bias categories. The fourth pie chart shows the overall distribution of tweets across bias categories. The fourth chart shows that in the dataset, 19.5\% of tweets are from outlets with left and right biases, 22\% and 21\% respectively from lean left and lean right, and 18\% from centrist outlets. The most interesting insights from this figure is that the tweets with neutral sentiments are posted exclusively by the centrist outlets. Tweets posted by journalists affiliated with left and right biased outlets tend to carry positive sentiments more often than negative sentiments while the reverse is found for those with lean left and lean right biases.  

\begin{figure}[h]
    \centering
    \includegraphics[width=\columnwidth]{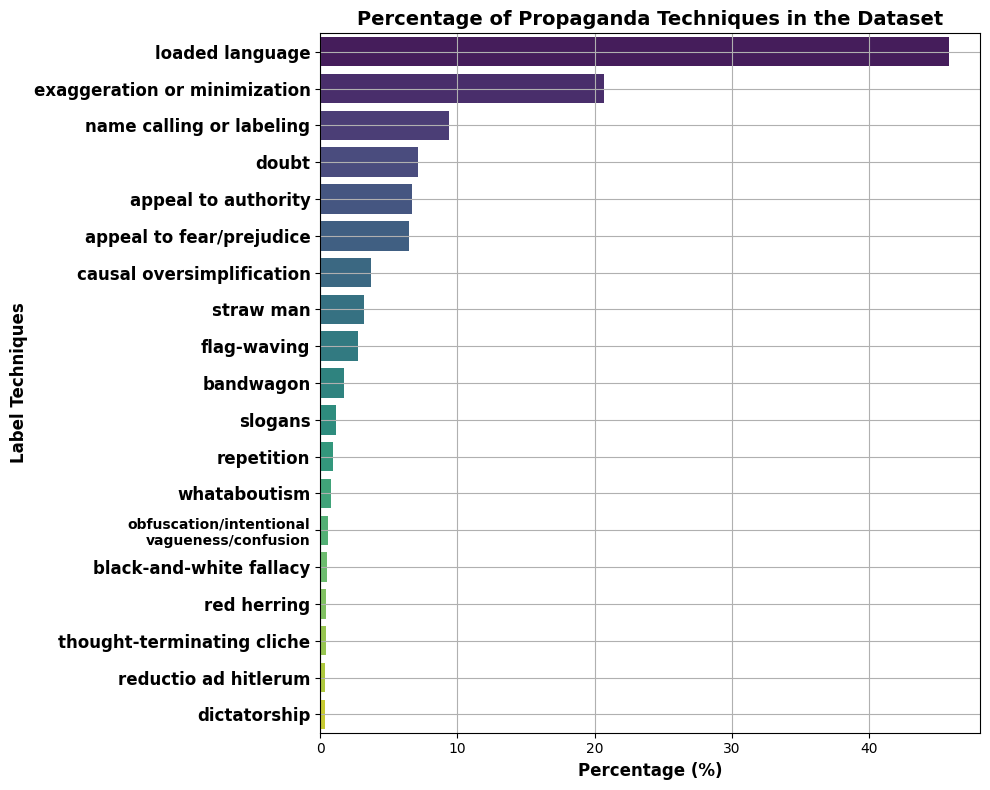}
    \caption{Propaganda types in annotated JMBX}
    \label{fig:freq_distribution}
\end{figure}

We present the frequency distribution of various propaganda techniques in the annotated dataset in Figure \ref{fig:freq_distribution}. The results indicate that loaded language is the most commonly used propaganda technique, appearing in 48\% of the tweets containing propaganda. Exaggeration or minimization is the second most common technique, found in 21\% of the tweets, while name calling or labeling is the third most frequent, present in 9\% of the tweets.

\section{\textbf{Experiments and Results}} 
\label{sec:exp_results}

We analyze a sample of 200 tweets to evaluate the performance of various LLM models using different prompts. The sample under analysis includes 100 propaganda and 100 non-propaganda tweets. The sample 100 propaganda tweets follow the propaganda type distribution in the corpus (see Figure~\ref{fig:freq_distribution}).

We run two sets of experiments. Within one set of experiments (Experiment 1) we evaluate the performance of the the trained BERT model on PTC and JMBX dataset. Within the other set of experiments (Experiment 2) we compare the outputs of eight LLMs on same datasets.

\subsection{Experiment 1: BERT model}
We use Purdue Anvil GPU system hosted by RCAC\footnote{\url{https://www.rcac.purdue.edu/}}~\cite{song2022anvil} utilizing a 3rd Gen AMD EPYC\textsuperscript{TM} 7763 CPU and NVIDIA A100 GPU for this experiment.
We fine-tune the pre-trained BERT uncased model\footnote{\url{https://www.kaggle.com/models}} on the PTC dataset~\cite{martino2020semeval}, which is widely used for detecting propaganda in news articles. The data is split with stratified technique for 70-10-20 train, validation, and test distributions.

\begin{table}[h]
\begin{center}
\begin{tabular}{ |c|c|c|c| } 
\hline
Dataset & Labels & P/R/F & F1(avg.)\\
\hline
\multirow{2}{1.5em}{PTC}&0 & 0.72/0.69/0.70 & \multirow{2}{1.5em}{0.71}\\ 
&1 & 0.70/0.73/0.71 & \\ 
\hline
\multirow{2}{2em}{JMBX}&0 & 0.60/0.84/0.70 & \multirow{2}{1.5em}{0.62}\\ 
& 1 & 0.73/0.40/0.54 &\\ 
\hline
\end{tabular}
\caption{Performance of the BERT-base-uncased model on PTC and JMBX dataset. 0 represents ``non propaganda'' while 1 represents ``propaganda''}
\label{table:bert_perf}
\end{center}
\end{table}

Table~\ref{table:bert_perf} contains the precision, recall, and F1-score for the classification task on both the PTC and JMBX datasets. With default hyperparameters, employing the Adam optimizer and binary cross-entropy as the loss function, the BERT-base model achieves an F1-score of 0.71 on the PTC test set. On the JMBX dataset, the model yields an F1-score of 0.62. Both experiments are conducted with the PTC dataset as the training set, utilizing 10 epochs and 7 different random seeds. Additionally, a default hyperparameterized RoBERTa model~\cite{liu2019roberta} is trained but it achieves a lower F1-score (0.66) compared to the BERT-base model.

\subsection{Experiment 2: Large Language Model}
We perform Zero-Shot prompting~\cite{radford2019language} on a sample of 200 tweets from the annotated dataset. Eight different Large Language Models from two most popular LLM providers, OpenAI and Google, are used. Five latest models from OpenAI are chosen namely: GPT 3.5 turbo, GPT 4, GPT 4 turbo, GPT 4o, GPT 4o-mini. Three models are selected from Google, Gemini 1 pro, Gemini 1.5 pro, Gemini 1.5 flash. As of August 2024, these are the latest models provided by these organizations. To the best of our knowledge, this is the first study that includes Google Gemini in the propaganda detection task while earlier studies were performed only on OpenAI's ChatGPT.  

\textbf{Experiment 2A:} The experiment relies on the Large Language Model's definition of propaganda. 

\textbf{Experiment 2B:} This experiment includes the definitions of 18 propaganda techniques used by~\cite{martino2020semeval} in the prompt. The LLM is asked to output ``propaganda'' if at least one of the propaganda technique is found in the tweet, otherwise the LLM outputs the ``non- propaganda'' label. The LLM can output ``not sure'' if it is unable to perform the classification.

\begin{table}[h]
\begin{center}
\begin{tabular}{ |c|p{1.5 cm}|c|c| } 
\hline
LLM & Model & 2A & 2B\\
& &P/R/F & P/R/F\\
\hline
\multirow{5}{3.5em}{GPT} & 3.5 & .70/.56/.61 & .75/.57/.63 \\
& 40613 & .74/.74/.74 & \textbf{.79/.79/.78} \\
&  4 turbo& .71/.69/.70 & .73/.72/.72 \\
&  4o & .79/.47/.59 & .79/.59/.67 \\
&  4o mini & .76/.56/.64 & .78/.69/.73\\
\hline
\multirow{3}{2.5em}{Gemini} & 1 pro & .68/.57/.62 & .71/.69/.69\\
& 1.5 Pro & .72/.54/.57 & .77/.56/.63 \\
& 1.5 Flash & .69/.59/.63 & \textbf{.70/.69/.69} \\
\hline
\end{tabular}
\caption{Performance of LLMs on PTC news dataset}
\label{table:exp2_ptc}
\end{center}
\end{table} 


\begin{table}[h]
\begin{center}
\begin{tabular}{ |c|p{1.5 cm}|c|c| } 
\hline
LLM & Model & 2A & 2B\\
& &P/R/F & P/R/F\\
\hline
\multirow{5}{3.5em}{GPT} & 3.5 & .68/.41/.51 & .71/.48/.57 \\
& 40613 & .74/.64/.60 & \textbf{.78/.71/.69} \\
&  4 turbo& .72/.61/.58 & .74/.66/.63 \\
&  4o & .80/.45/.53 & .80/.56/.64 \\
&  4o mini & .72/.50/.53 & .76/.61/.63 \\
\hline
\multirow{3}{2em}{Gemini} & 1 pro & .71/.53/.54 & .71/.64/.61\\
& 1.5 Pro & .70/.52/.59 & .79/.51/.61 \\
& 1.5 Flash & .69/.57/.62 & \textbf{.74/.68/.67} \\
\hline
\end{tabular}
\caption{Performance of LLMs on JMBX dataset}
\label{table:exp2_jmbx}
\end{center}
\end{table}
\vspace{-0pt}

\begin{figure}[h]
    \centering
    \includegraphics[width=\columnwidth]{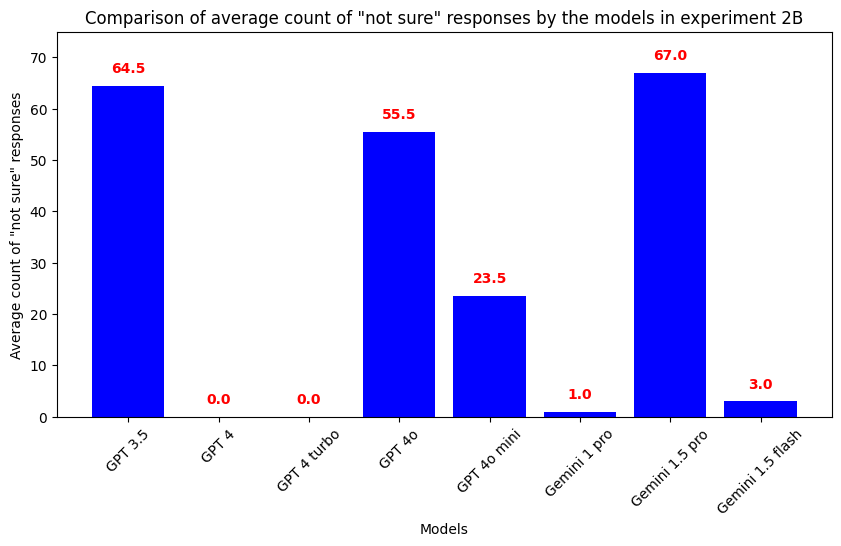}
    \caption{Number of "not sure" output by various LLMs on experiment 2B.}
    \label{fig:not_sure}
\end{figure}

Figure~\ref{fig:not_sure} shows the average number of ``not sure'' returned by the models in experiment 2B. Results presented in Table~\ref{table:bert_perf} show that adding definition of propaganda increases the performance on all LLMs.

\vspace{-0pt} 

\textbf{Experiment 2C:} In experiment 2C, we use Chain of Thought (CoT)~\cite{wei2022chain} prompting technique. The LLMs are prompted with the definition of propaganda techniques similar to Experiment 2B with an additional ``think step by step'' type strategy at the end. We select two best performing models from each organization that produced fewer instances of ``not sure'' classifications when categorizing tweets. As shown in Table~\ref{table:exp2_ptc_jmbx}, the recall and F-score increased on the JMBX dataset with the use of CoT prompting. However, this improvement is not observed in the PTC dataset. Notably, for the Gemini 1.5 Flash model, CoT prompting yields improved performance on both the datasets.

\begin{table}[h]
\begin{center}
\begin{tabular}{ |c|c|c| } 
\hline
Model & PTC & JMBX\\
\hline
GPT 4 0613 & .77/.77/.76 & .75/.73/.73 \\
Gemini 1.5 flash & .76/.73/.72 & .73/.73/.73 \\
\hline
\end{tabular}
\caption{Results of Experiment 2C on GPT 4 and Gemini 1.5 models}
\label{table:exp2_ptc_jmbx}
\end{center}
\end{table}
\vspace{-5pt} 

\section{Environmental Impact}
\label{sec:env}
We estimate the carbon footprint of this research and of any product that utilizes the techniques we propose. This research tasks involves classifying 200 tweets with prompts to LLM, averaging runtime to 4 minutes per task. Assuming the task ran on GPT-3/4 models hosted on Microsoft Azure servers (utilizing a single Nvidia A100 GPU), for a total of 2 hours (4 minutes per model, 5 models, 3 runs, 2 datasets), the estimated carbon emission was 0.28 kg, based on the mlCO2 calculator by Lacoste et.al.~\cite{lacoste2019quantifying}. Similarly, for the Gemini model, hosted on Google Cloud Platform, with a runtime of 1.2 hours (4 minutes per model, 3 models, 3 runs, 2 datasets), the estimated carbon emission was 0.11 kg. The total emission for this research is 0.39 kg.

Since the mlCO2 calculator does not factor in the Power Usage Effectiveness (PUE) of data centers, we apply an average PUE of 1.12, based on recent studies~\cite{faiz2023llmcarbon}, and reports from Google~\cite{google_datacenter} and Microsoft~\cite{ms_datacenter}, which adjusts the total carbon emission to 0.44 kg. 
While this value might seem negligible, according to the EPA calculator~\cite{EPA_calculator}, it is roughly equivalent to driving a car for a mile, or charging 29 smartphones.

\section{\textbf{Conclusion \& Future work}}
\label{sec:conc}
We present an annotated dataset containing tweets posted by highly followed journalists and insights into the relationship between the propaganda in their tweets and the political bias of their affiliation. We fine-tune a BERT base model on a well-known propaganda dataset, as well as use zero-shot and CoT LLM prompting techniques to measure propaganda detection by eight LLM models on these datasets. Our primary insight in this work is that journalists affiliated with extremely biased news outlets tend to use more propaganda in their writings than those affiliated with moderately biased organizations. We also report that zero-shot prompting on LLMs shows better performance in detecting propaganda than the BERT model, and with Chain of Thought prompting, the performance is further improved. We estimate the environmental impact of using LLMs in this research. Future research will focus on evaluating the performance of large language models in the fine-grained detection of individual propaganda techniques.

\section{\textbf{Limitations \& Discussion} }

We recognize that there are limitations to this research due to various issues related to data collection and data sharing. Even with paid subscription, the number of tweets one can retrieve using the API is limited. Additionally, the tweets classified under the "center" category are sourced from official media accounts, which complicates direct comparison with the personalized tweets from journalists associated with biased media outlets. Another limitation involves the environmental impact estimates. These estimates are derived from related studies, online reports, and publicly available tools designed to calculate emissions and energy consumption. However, these sources acknowledge a degree of inaccuracy in their results, as various factors, such as data center efficiency and hardware specifications, can influence the final calculations.

\section{\textbf{Ethical Considerations} }
This study addresses several ethical considerations especially when dealing with data from social media platforms. To mitigate potential privacy violations, we anonymized all user data and ensured that personally identifiable information was not included by redacting journalist's name, their affiliation, their meta information in the dataset. Additionally, the potential biases due to selection of certain journalists in data must be acknowledged. To address this, we ensured a balanced representation of perspectives within the dataset. However, we recognize that no dataset or model is entirely free of bias, and thus our findings should be interpreted with caution. Another critical aspect is the ethical implications of propaganda detection. Identifying and labeling content as propaganda could have significant societal impacts, including influencing public perception and discourse. Therefore, we included subject matter experts to understand nuanced interpretations of the content. Furthermore, we emphasize transparency in the model's decision-making process and make the dataset publicly available, ensuring that the detection results can be replicated, scrutinized and understood by stakeholders. Finally, energy consumption and environmental impact were considered throughout the research. Training LLMs are computationally intensive and contributes to carbon emissions ~\cite{strubell2020energy}. We aimed to mitigate this by being aware and optimal use of Large Language Model and sampling a subset of dataset in our experiments. Future research should continue exploring greener alternatives for model training and deployment.

\printbibliography

\end{document}